\documentclass[11pt,twoside]{article}

\pdfoutput=1

\usepackage{asp2006}
\usepackage{epsf}
\usepackage{graphicx}
\usepackage{lscape}
\usepackage{amsbsy}
\usepackage{amssymb}

\begin{document}
%%% Fill in title
\title{The ground-based solar observations database BASS 2000}

%%% Fill in author names
\author{F. Paletou, M. Lafon, P. Maeght, F. Grimaud and T. Louge}
\affil{Universit\'e de Toulouse, Observatoire Midi-Pyr\'en\'ees, LATT
  (CNRS/UMR5572), 14 ave. E. Belin, F-31400 Toulouse, France}

\author{J. Aboudarham} \affil{Observatoire de Paris, LESIA
(CNRS/UMR8109), 5 place J. Janssen, F-92195 Meudon, France}

\begin{abstract} 
BASS 2000 is the French solar database for ground-based
instruments. We describe hereafter our organization, our tasks and the
products we can deliver to the international community. Our prospects
cover data mining into the TH\'eMIS archive, a participation to the
EST endeavour and the creation and curation of the ESPaDOnS/NARVAL
stellar spectra database.
\end{abstract}

%%% Top level section head (remove "%" symbol)
\section{Organization, tasks and products}   
BASS 2000\footnote{{\tt http://bass2000.bagn.obs-mip.fr}} is the
archive and database of several ground-based solar instruments such as
the NRH (Nan\c cay Radio-Heliograph), the \emph{Observatoire de
Paris-Meudon} spectroheliograph, the \emph{Pic du Midi} coronograph
(HACO, now CLIMSO) and the French-Italian telescope TH\'eMIS installed
in Tenerife.

BASS 2000 is located at two different sites. Its main centre is hosted
by the \emph{Observatoire Midi-Pyr\'en\'ees}, and it is situated on
the campus of Tarbes about 150 km south-west of the main town of
Toulouse. It is the main database and archive of TH\'eMIS. The later
archive represents now about 11 To of mostly raw data collected since
1999.

Concerning TH\'eMIS data, it is quite important to recall its data
policy here. Indeed, after a PI-ship of one year, \emph{all data
become public}. On-line query forms are available to users in order to
select data of interest for them. Afterwards we take care of
supplying to any user the requested data over the network or using
standard media, depending on the requested volume, as quickly as
possible.

The BASS 2000 centre at \emph{Observatoire de Paris-Meudon} is mainly
in charge of systematic data from other observatories than TH\'eMIS.
Daily images of the Meudon spectroheliograph, of the Nan\c cay
decametric array and radio-heliograph, and of the \emph{Pic du Midi}
coronograph are indeed available here. This centre also provides
additional services and tools such as solar ephemeris and reference
spectra on-line.

Finally, a service aka. FROMAGE in charge of computing, upon request,
extrapolations of photospheric vector magnetic field maps to the
corona is now in operations.

Since 2006, vector magnetic field maps are available on-line, when
pointing at the Tarbes archive's web-page (see Fig. 1). For the sake of
traceability, users will be gently asked to be identified before
accessing to the data. Inversion of data are made using either a
PCA-based code \citep[][]{pca} or a Levenberg-Marquardt code
\citep[][]{unnofit}.

\begin{figure}
  \centering
  \includegraphics[width=14cm,angle=0]{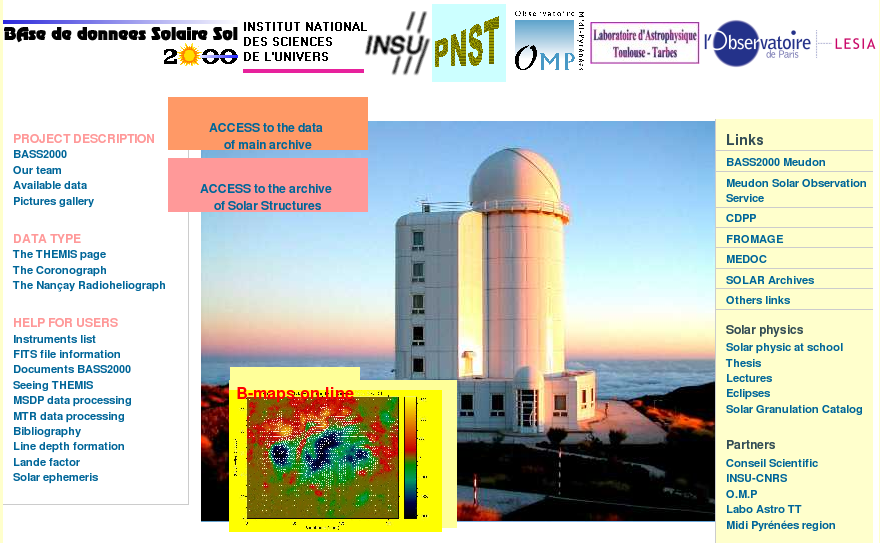}
  \caption{Outlook of the Tarbes archive's web-page of BASS
  2000. There is a direct link to some of the reduced date on-line
  consisting, so far, in vector magnetic field maps.}
%%  \label{Fig1}
\end{figure}

In 2007, automatic reduction and data transfer procedures have been
succesfully tested between TH\'eMIS and BASS 2000. This has been done
during a 3-week campagn together with HINODE, last fall.

\section{Prospects}
The experience of the relation between TH\'eMIS, as a data producer,
and BASS 2000, its main archive and associated data base is unique in
the solar physics community.

On a short/medium-term basis, BASS 2000 will aim at providing the best
scientific return from TH\'eMIS data. We shall therefore carry-on
\emph{upon request} the backward reduction of MTR \citep[][]{fpgm}
data taken before 2007, using the standard reduction package $SQUV$
provided by the TH\'eMIS group \citep[][]{squv}.

Beyond the production of vector magnetic field maps, plans are now for
allowing data mining in the large set of second solar spectrum for
atomic and molecular lines data, for instance.  This is also planned
for new solar prominences data taken simultaneously in the visible and
near-infrared spectral ranges \citep[see e.g.,][]{llfpspw5}.

The BASS 2000 group is involved in the \emph{European Solar Telescope}
design study which will be funded by the European Union's FP7. Our
main contribution into the data flow system workpackage will be the
survey of the development of Virtual Observatory standards and
tools.

Meanwhile the design and the implementation of EST, another
evolution of BASS 2000 could be the integration of several other
ground-based data coming from other European instruments/telescopes.

Finally, we should open early in 2008 a database for
ESPaDOnS\footnote{{\tt
http://www.ast.obs-mip.fr/projets/espadons/espadons.html}} and NARVAL
stellar spectropolarimetric data. The first release should deal with
$\sim$ 1\,000 spectra taken since 2005 both at the CFHT and TBL
telescopes mainly on F, G and K-type stars.

\acknowledgements Our warmest thanks go to Nad\`ege Meunier, head of
BASS 2000 during the last 6 years. We are indebted to the French
\emph{Programme National Soleil-Terre} and to the \emph{Universit\'e
Paul Sabatier, Toulouse 3} for their substantial financial support.


\begin{thebibliography}{}
\bibitem[Bommier et al.(2007)]{unnofit}
Bommier, V., Landi Degl'Innocenti, E., Landolfi, M. \& Molodij, G. 2007, \aap, 464, 323
\bibitem[L\'eger \& Paletou(2008)]{llfpspw5}
L\'eger, L. \& Paletou, F. 2008, these proceedings
\bibitem[Paletou \& Molodij(2001)]{fpgm}
Paletou, F., \& Molodij, G. 2001, in ASP Conf. Ser. 236,
Advanced Solar Polarimetry, ed. M. Sigwarth, (San Francisco: ASP), 9
\bibitem[Rees et al.(2000)]{pca}
Rees, D., L\'opez Ariste, A., Thatcher, J. \& Semel, M. 2000, \aap, 355, 759 
\bibitem[Sainz Dalda \& L\'opez Ariste(2007)]{squv}
Sainz Dalda, A. \& L\'opez Ariste, A. 2007, \aap, 469, 721 
\end{thebibliography}
\end{document}